# Pore Scale Study of Gas Bubble Nucleation and Migration in Porous Media

Nariman Mahabadi[1] & Leon A. van Paassen[2]
[1]*University of Akron, Department of Civil Engineering*
[2]*Arizona State University, School for Sustainable Engineering and the Built Environment*

ABSTRACT: The behavior of gas bubbles in porous media includes bubble nucleation and growth, migration, coalescence, and trapping. These processes are significantly affected by the pore scale characteristics and heterogeneity of the sediment. In this study, experiments are performed using a microfluidic chip in which different gas bubble behavior mechanisms are observed. Based on the microscale experiments, eight major gas bubble behavior mechanisms are identified. In addition, a mesoscale experiment is designed and performed to study the impacts of soil layering and heterogeneity on the formation and distribution of biogenic gas generation via denitrification. The results show that the pore scale characteristics of soil significantly affect the distribution and migration pattern of biogenic gas bubbles. As a result, the water saturation may vary locally between fully saturated (in fine sand), to about 80% in zones where the gas is allowed to migrate freely (coarse sand), to even close to 0% in the zone where the gas gets trapped under layers with higher air entry value.

## 1 INTRODUCTION

Subsurface sediments can be de-saturated by various natural gas formation processes such as microbial activity in shallow ocean sediments or wetlands (Whalen, 2005; Leifer & Patro, 2002; Abrams, 2005), air trapping by groundwater-level oscillation (*Krol et al.*, 2011), changes in atmospheric pressure, gas solubility change due to seasonal temperature variation in the subsurface (*Ryan et al.*, 2000) methanogenic degradation of hydrocarbon contaminants in the subsurface (*Amos et al.*, 2005), and decomposition of municipal solid waste in landfills (*van Breukelen et al.*, 2003). In addition, the leakage-induced depressurization carbon dioxide dissolved brine during the long-term geological $CO_2$ sequestration may results in gaseous $CO_2$ formation in the subsurface (*Plampin et al.*, 2014; *Zuo et al.*, 2012; *Zuo et al.*, 2013).

On the other hand, gas bubbles can also be introduced in the subsurface artificially and have been used to remediate contaminated soils, enhance the extraction of resources or alter the hydrological or mechanical properties of subsurface sediments. Gas injection or in situ gas formation has several applications such as liquefaction mitigation by microbially induced desaturation via denitrification (*He and Chu*, 2014; *Rebata-Landa and Santamarina*, 2012; O'Donnell et al., 2017, van Paassen et al., 2017), gas exsolution or air sparging by supersaturated water injection (SWI) for groundwater/soil remediation (*Enouy et al.*, 2011; *McCray and Falta*, 1997), viscosity reduction by heavy oil depressurization (*Bora et al.*, 2000; *Stewart et al.*, 1954), methane gas production from hydrate-bearing sediments (*Jang and Santamarina*, 2011, 2014; Mahabadi and Jang 2014, Mahabadi et al., 2016a, 2016b), and $CO_2$ sequestration/$CO_2$ foam injection (*Zheng and Jang*, 2016; *Zheng et al.*, 2017). The gas formation mechanisms in the abovementioned applications include nitrate reduction, direct gas bubble injection, depressurization, temperature increase, electrolysis, and drainage-recharge.

Once the gas bubbles are formed in the sediment, they can migrate upward due to the buoyancy, or are sometimes trapped in the pore space. The gas nucleation, migration, and trapping and the associated impacts are frequently found in the in-situ sediment. Methane ebullition, the release of methane into the atmosphere or the movement through porous media, is the typical mechanism of greenhouse gas emission from aquatic ecosystems (*Amos and Mayer*, 2006; *Ramirez et al.*, 2015; *Walter et al.*, 2006). The gas bubble formation also affects the mechanical properties of the sediment (*Grozic et al.*, 1999; *Sills et al.*, 1991). Due to gas compressibility, even a small fraction of gas is sufficient to reduce pore fluid bulk stiffness (Biot, 1941; Skempton, 1956) and dampen pore pressure build up during monotonic and cyclic undrained loading (Yang et al., 2004; Yegian et al., 2007; He and Chu 2014). However, when a large fraction of gas gets trapped forming gas pockets, the upward buoyancy force counteracts the overburden pressure, reducing the effective stress, and may generate fractures and lifts up the overlying soil (van Paassen et al., 2017). In addition, very small gas bubbles trapped in the porous media can dramatically reduce hydraulic conductivity without the significant reduction in water saturation (*Ronen et al.*, 1989; Mahabadi et al., 2017).

The behavior of gas bubbles in porous media includes bubble nucleation and growth, coalescence, migration and trapping. All these mechanisms are significantly affected by the pore scale characteristics and heterogeneity of the soil. In this study, a microscale experiment is performed using a microfluidic chip in which several gas bubble behavior mechanisms are detected and explained in detail. In addition, a mesoscale experiment is designed and conducted to study the impacts of soil layering and heterogeneity on the formation and distribution of biogenic gas generation via denitrification in soil.

## 2 EXPERIMENTAL STUDY

### 2.1 Microscale experiment

A Two-dimensional transparent microfluidic chip (MICRONIT Microfluidics, Netherlands) is used, which was designed and fabricated to resemble a homogenized particle packing. The dimensions of the microfluidic chip is 21.3mm×12.7mm, where the internal thickness (pore depth) is 50 µm. The microfluidic chip includes 377 circular mono-sized grains with 800µm of dimeter and the size of pore throat between two grains is 140µm (Figure 1a).

Figure 1b shows the configuration of the microfluidic chip setup. The microfluidic chip is placed vertically while it is fixed in a steel-frame holder.

In this study, a commercial dental product called Efferdent is used as the source of gas bubbles. Efferdent is an antibacterial dental cleanser which rapidly dissolves in water and can be used to generate micro-sized oxygen bubbles within porous media. The main ingredient of Efferdent is sodium perborate which is an active oxygen bubble source.

In laboratory scale experiments, Efferdent has advantages over other sources for gas generation which is mainly due to its fast reaction with water, the ability to control the degree of saturation, and uniformity of gas bubble distribution throughout the pore space (Eseller-Bayat 2009).

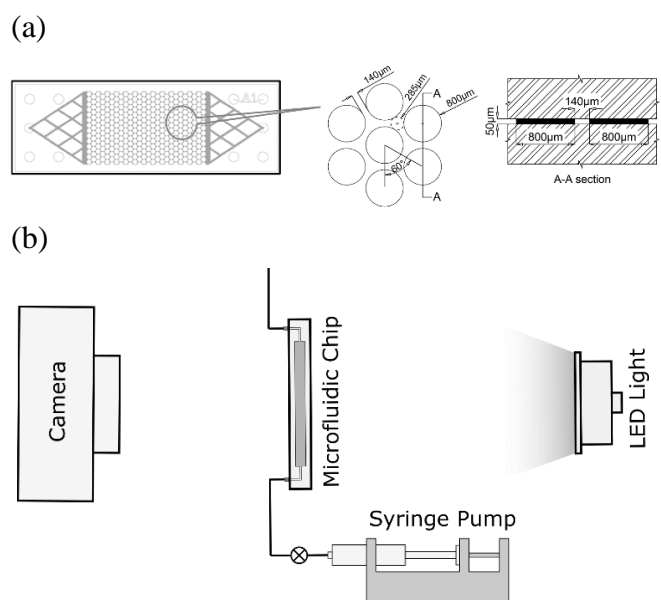

Figure 1. (a) Experimental configuration for the microfluidic chip test; (b) Geometry of the patterned microfluidic chip

In order to generate the oxygen bubbles throughout the pores of microfluidic chip, 1 wt % Efferdent is mixed with 99 wt% cold water (1ºC). The prepared solution is injected to the microchip using a syringe pump connected to inlet port of the chip. Since Efferdent-water reaction is extremely fast (less than a second), decreasing the temperature of water allows delayed reaction in which the reactive solution can be injected to the microfluidic chip before generating the gas bubbles.

After the injection, the inlet port is closed while the outlet port is remained open which allows movements of buoyancy driven gas bubbles.

Once the injection is completed. The behavior of gas bubbles is recorded using a digital camera (D5200 Nikon equipped with a 60 mm micro lens AF-S NIKKOR f/2.8G).

Figure 2 presents eight different gas bubble mechanisms including nucleation, expansion and migration which are detected using the results of the microfluidic chip experiment. A detailed illustration for each mechanism is provided herein:

(a) *Nucleation and Expansion:* Oxygen gas generated by the reaction between Efferdent and water initially remains dissolved as long as the concentration is lower than equilibrium as defined by Henry's law:

$$K_H = \frac{c_g}{p_g}$$

where $K_H$ is Henry's constant, $c_g$ is the dissolved gas concentration in the liquid phase and $p_g$ is the partial pressure in the gas phase. When the concentration reaches a sufficiently high supersaturation condition, gas bubbles may nucleate. The molecular interactions between dissolved gas and the liquid defines the supersaturation threshold for homogeneous nucleation. However, the presence of mineral cavities, impurities, and irregularities tends to favor heterogeneous gas bubble nucleation even at lower supersaturation conditions (Blandar 1979, Gerth and Hemmingsen 1980; Pease and Blinks 1947; Dominguez et al. 2000; Rebata-Landa and Santamarina 2012). The nano-sized gas bubbles that formed at the favorable nucleation sites are in stable condition only if the critical size is reached. Gas bubbles smaller than the critical size re-dissolve into the liquid as the gas partial pressure, which is a function of the bubble radius, R, surface tension, $\tau$ (0.072 N/m for water at 20ºC) and the pore water pressure, $P_w$:

$$P_g = P_c - P_w = \frac{2\tau}{R} - P_w$$

After nucleation, small bubbles may still re-dissolve, diffuse, and agglomerate into the larger bubbles, through a phenomenon called Ostwald-ripening (Schmelzer and Schweitzer, 1987). The diffusion flux of the dissolved gas molecules results in the growth of the pre-existing larger gas bubbles. The growing gas bubbles continue to expand, while the pressure decreases, until it reaches the neighboring pore throats. At this moment, the gas bubble gets stuck due to the capillary pressure between the gas bubble surface and the water phase in the pore throat. Now the capillary pressure $P_c$, is expressed as:

$$P_c = \frac{2\tau}{r}$$

In which $r$ is the radius of pore throat. The continuous diffusion of the dissolved gas towards the stuck bubble results in the increase of internal gas bubble pressure. Once the gas bubble pressure exceeds the capillary pressure, the stuck bubble expands to the neighboring pore. Once the air entry value of the pore is exceeded, the gas bubble squeezes through the pore throat and expands to the neighboring pore, while the gas volume increases and the gas pressure drops. This uneven dynamic displacement of gas bubbles is often referred to as Haines jumps (Armstrong et al. 2015).

(b) *Upward movement towards larger pores:* Once the gas bubbles are generated in the pore space, they can migrate in upward direction due to their buoyancy force. However, the restriction of pore throats limits their movement. In this condition, the moving bubble chooses the largest pore throat in which the associated capillary pressure (air entry value) is minimum.

(c) *Upward movement towards blocked pores:* The largest pore throat is not always the preferred choice for the moving bubble. Sometimes the adjacent upper-level pore to the largest pore throat is already occupied by another gas bubble. In this condition, the moving bubble changes its direction and picks the second larger pore throat where there is a free percolation path towards the outlet.

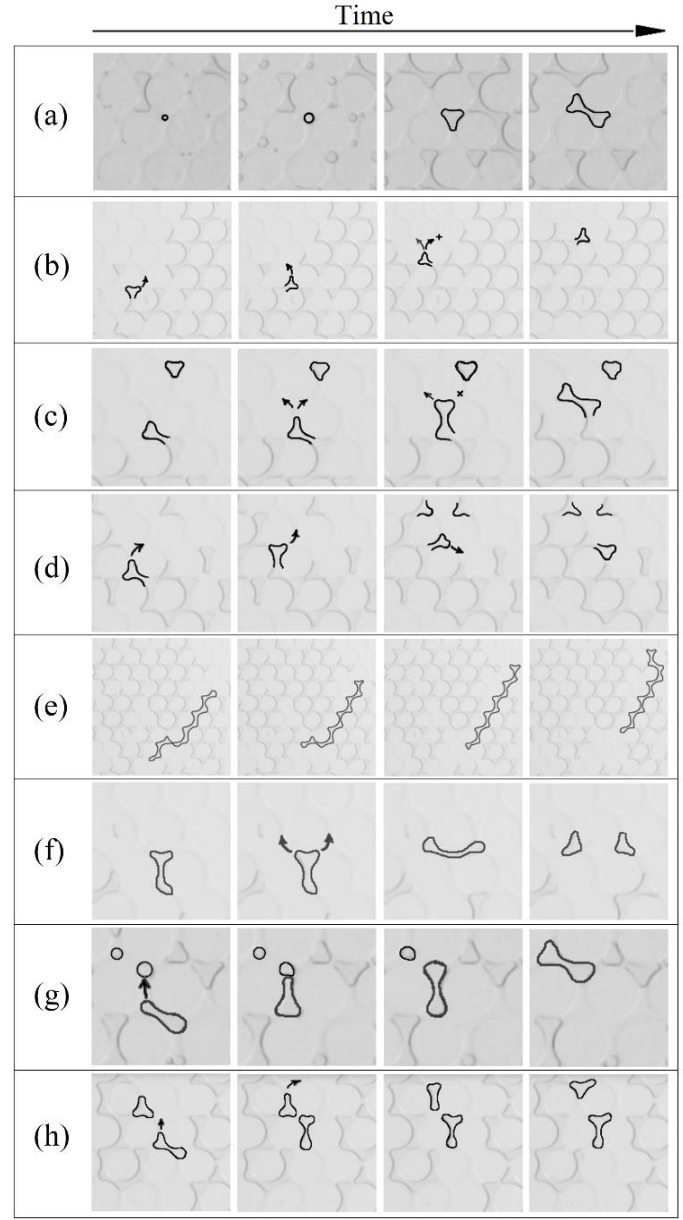

Figure 2. Various mechanisms observed during the gas bubble formation and migration in microfluidic chip. (a) Nucleation and expansion. (b) Upward movement towards larger pores. (c) Upward movement towards blocked pores. (d) Backward movement. (e) Slug movement. (f) snap-off. (g) Coalescence. (h) Pushing.

(d) *Backward movement:* Generally, the gas bubbles tend to move upward due to their buoyancy. However, if all the upper-level pores are trapped by pre-existing bubbles, the moving gas bubble seek all the possible pathways towards the outlet boundaries, even those located in lower-level depths.

(e) *Slug movement:* All the previous movement mechanisms (a, b, c and d) involve a series of dynamic actions resulted from the struggle between the internal gas pressure increase and buoyancy as the driven force, and the capillary pressure and internal gas pressure release as the resistive component. However, if the gas bubble is long enough, the buoyant force is large enough in which the gas bubble can freely squeeze through the pore throats and move like a slug [Roosevelt & Corapciaglu, 1998]. The formation of slug movements is affected by the pore scale characteristics of the porous media (pore size, connectivity and heterogeneity).

(f) *Snap-off:* The snap-off mechanism occurs when the interface water in the corner layers in a throat swell until it is in a state pressure non-equilibrium resulting in spontaneous filling of the pore throat and disconnection of the gas bubble (Singh et al., 2017). During snap-off, water accumulates in the vicinity of throats due to gradients in the axial component of the curvature of the water-gas interface. Snap-off mechanism is a function of pore geometry and wettability (Wardlaw 1982).

(g) *Coalescence:* During migration, gas bubbles may coalesce and form bigger gas bubbles. This mechanism is not always effortless since the trapped water between two adjacent gas bubbles must be expelled. If the trapped gas bubble is smaller than the hosting pore, there would be a gap between the gas bubble and surrounding soil particles, so the trapped water can squeezed out and upcoming gas bubble can coalesce with the trapped bubble, making a bigger bubble.

(h) *Pushing:* The upward movement of gas bubbles may affect the stability of pre-existing gas bubbles which located in upper depths. Pushing mechanism occurs when the buoyancy-driven movement of bubbles results in an increase in the velocity field of water in the vicinity of the gas bubble front. Sometimes the resulted velocity change is large enough to push the pre-existing bubbles and support them in order to be released from the hosting pore.

## 2.2 Mesoscale experiment

An experimental cell is designed to study the formation and migration behavior of biogenic gas in a heterogeneous soil. The cell consists of two transparent acrylic plates where the thicker plate has a pocket to house the soil. The dimensions of the pocket inside the transparent cell is 22.9 cm × 18.8 × 0.94 cm. The pocket is filled with Ottawa 20-30 (coarse sand, $D_{50}$=0.72 mm) and Ottawa F60 (fine sand, $D_{50}$=0.21 mm). Figure 3 shows the configuration of the experiment and the layering of sands. Once the pocket is filled with sand it is oversaturated with water. The water level is set to be higher than the soil layers in order to allow measurements of the amount of gas generation during the test. A solution is prepared to stimulate biogenic nitrogen gas formation by nitrate reducing bacteria, containing 12 mM of calcium acetate, 10mM of calcium nitrate and 0.5 mL/L of trace minerals and inoculated with a mixed bacterial culture (Pham et al 2016, 2018). The solution is dyed with a food color liquid (blue) in order to facilitate the image processing and identify the migration of injected liquid. The denitrifying solution (30 mL) is injected using a syringe from the inlet port located at the bottom of the cell.

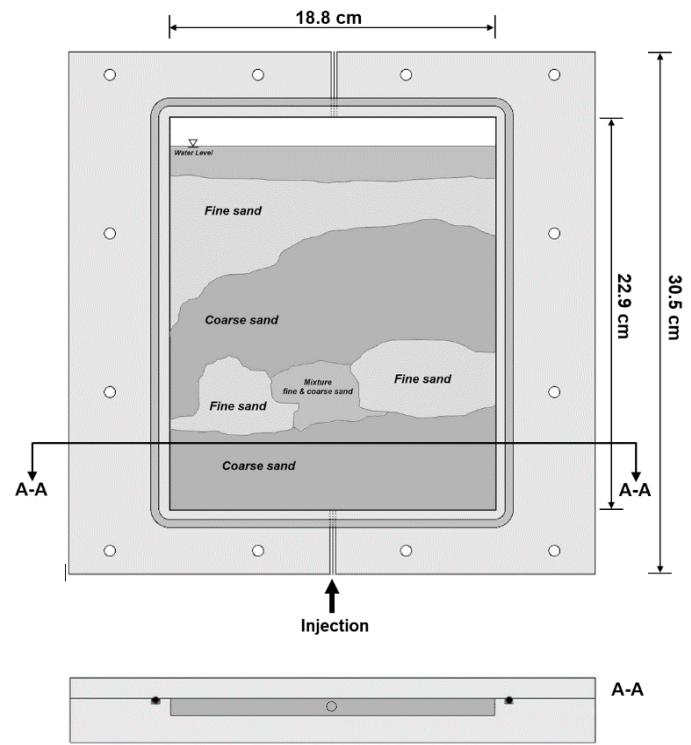

Figure 3. Experimental configuration for mesoscale study. A transparent cell to study biogenic gas formation and migration in soil.

The injected denitrifying solution almost filled the coarse layer located at the bottom of the cell completely. Once the injection is completed, the inlet is closed while the outlet remained open.

Using a digital camera, interval images are taken every 10 minutes for 17 days. Imaging is stopped once no more gas generation is observed. An image processing technique is developed to analyze the images and detect the location of biogenic gas bubbles, and measure the amount of generated gas.

The first image (at the beginning of the test, right after the injection) is set as the benchmark, while all the images taken later are compared to this image. Using an image correlation technique, the location of gas bubbles is detected. For each image, the rise in water level is used to calculate the volume of generated gas generation and the degree of saturation.

Figure 4 shows the image processing results for 3, 5, 7, 9, 11 and 13 days after the injection. The results show 3 days after the injection of denitrifying solution, some gas bubbles are formed close to the injection point. It should be noted that due to the injection of substrate, a small cavity is formed close to the injection port which then found to be a preferable nucleation zone for gas bubbles.

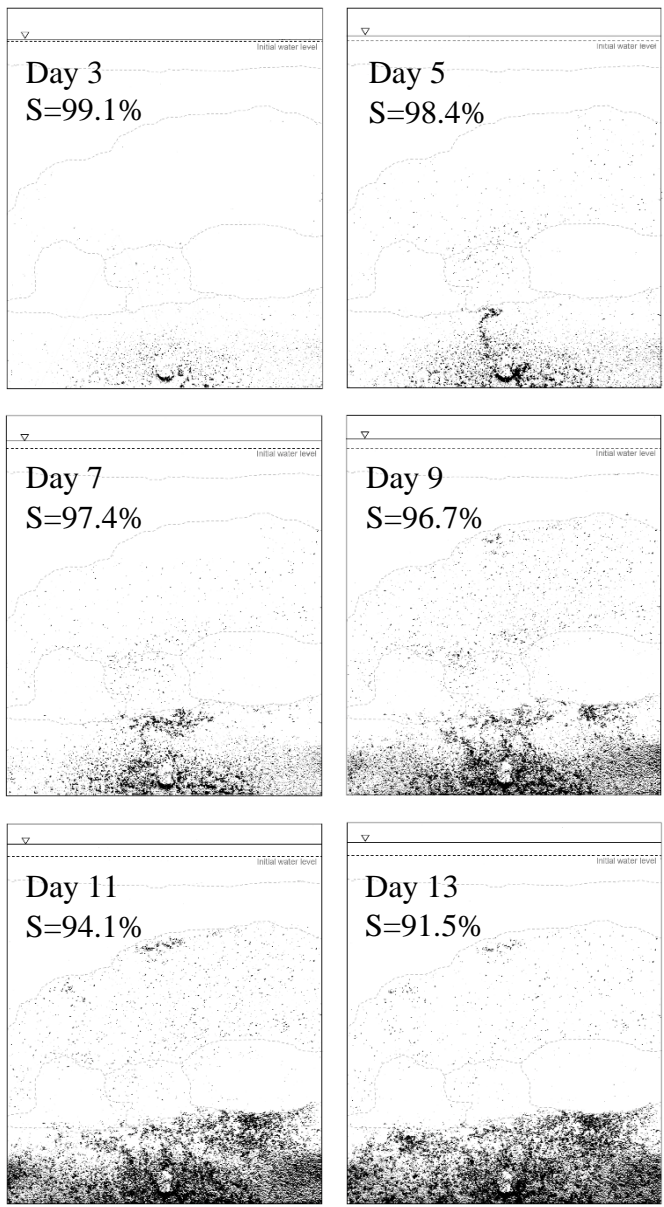

Figure 4. Gas distribution and the degree of saturation for 3, 5, 7, 9, 11, and 13 days after injection of denitrifying solution.

In day 5, more gas bubbles are formed and migrated towards the top layer. Since the second layer (from bottom) is a fine sand layer, due to the high capillary pressure, the migrated gas bubbles were not be able to percolate through this layer, and turned to the right side of the screen in order to find an easier pathway. The results of day 5 show that only a few gas bubbles have passed the mixture of fine-coarse layer (located between the two fine sand layers) and reached to the third layer (coarse sand layer).

In days 7 to 13 more gas is formed and a few gas bubbles were able to reach to the boundary between the fine and coarse layer close to the top of the screen. The average degree of saturation is decreased to 91.5% at day 13, after which no further gas generation is observed. No percolation of gas bubbles is observed through the topmost layer which indicates that the total volume change of water (water level change) can be considered as the total volume of gas generation and the trapped gas is reasonably stable. The total volume change due to gas formation is 10 mL at the end of the test. The gas migration in the heterogeneous soil is affected by the pore scale characteristics. Gas migration through the fine sand layer is limited due to the high air entry value required for gas bubbles in order to squeeze through the pore space. In order to identify the air entry value for the soils used in this study, the soil water characteristic curves of Ottawa 20-30 and Ottawa F60 are measured using a HYPROP device (Figure 5). The air entry value is measured around 1 kPa for Ottawa 20-30, and 3 kPa for Ottawa F60. The residual water contents for Ottawa 20-30 and Ottawa F60, i.e. the remaining water content at high suction, are 2 and 5% respectively.

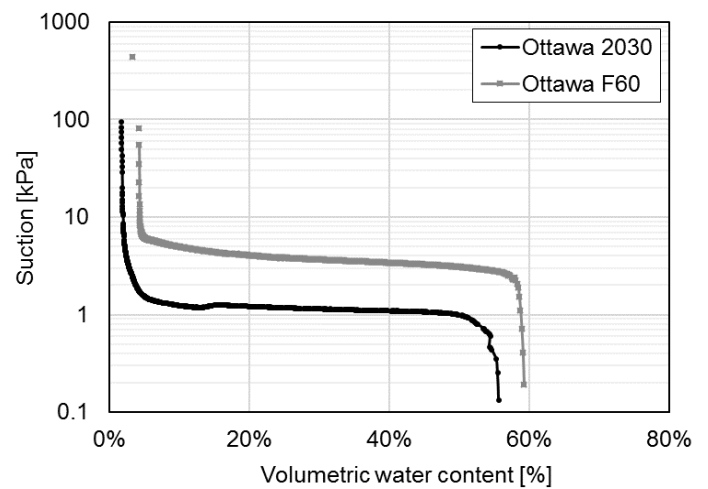

Figure 5. Soil water characteristic curve for coarse sand (Ottawa 20-30) and fine sand (Ottawa F60).

## 3 CONCLUSION

The behavior of gas bubbles in porous media includes bubble nucleation and growth, migration, coalescence, and trapping. These processes are significantly affected by the pore scale characteristics and heterogeneity of the sediment. In this study, experiments are performed using a microfluidic chip in which different gas bubble behavior mechanisms are observed. Based on the microscale experiments, eight major gas bubble behavior mechanisms are identified. In addition, a mesoscale experiment is designed and performed to study the impacts of soil layering and heterogeneity on the formation and distribution of biogenic gas generation via denitrification. The results show that the pore scale characteristics of soil significantly affect the distribution and migration pattern of biogenic gas bubbles. As a result, the water saturation may vary locally between fully saturated (in fine sand), to about 80% in zones where the gas is allowed to migrate freely (coarse sand), to values close to the residual water content even close to the residual water content in zones where the gas gets trapped under layers with higher air entry value.